\documentclass[final,5p,twocolumn]{elsarticle}
\usepackage{amsthm,amsmath,amssymb,graphicx,dcolumn,bm,hyperref}
\usepackage{soul}
\usepackage{color}
\journal{Journal of Magnetism and Magnetic Materials}

\begin{document}

\begin{frontmatter}

\title{Theory of Harmonic Hall Responses of Spin-Torque Driven Antiferromagnets}

\author[inst1]{Hantao Zhang}
\ead{hzhan289@ucr.edu}
\author[inst1,inst2]{Ran Cheng}
\ead{ran.cheng@ucr.edu}

\affiliation[inst1]{organization={Department of Electrical and Computer Engineering, University of California},
            addressline={900 University Avenue}, 
            city={Riverside},
            postcode={92521}, 
            state={California},
            country={United States}}
\affiliation[inst2]{organization={Department of Physics and Astronomy, University of California},
            addressline={900 University Avenue}, 
            city={Riverside},
            postcode={92521}, 
            state={California},
            country={United States}}

\begin{abstract}
Harmonic analysis is a powerful tool to characterize and quantify current-induced torques acting on magnetic materials, but so far it remains an open question in studying antiferromagnets. Here we formulate a general theory of harmonic Hall responses of collinear antiferromagnets driven by current-induced torques including both field-like and damping-like components. By scanning a magnetic field of variable strength in three orthogonal planes, we are able to distinguish the contributions from field-like torque, damping-like torque, and concomitant thermal effects by analyzing the second harmonic signals in the Hall voltage. The analytical expressions of the first and second harmonics as functions of the magnetic field direction and strength are confirmed by numerical simulations with good agreement. We demonstrate our predictions in two prototype antiferromagnets, $\alpha-$Fe$_{2}$O$_{3}$ and NiO, providing direct and general guidance to current and future experiments.
\end{abstract}


\end{frontmatter}

\section{Introduction}

To meet the growing demand for faster, smaller and more energy efficient magnetic devices, antiferromagnetic (AFM) spintronics emerges as a new area of active research~\cite{BaltzReview2018,GomonayReview,Jungwirth2018}. Owing to the vanishing stray fields and ultrafast dynamics enabled by the strong exchange interactions, AFM materials hold huge potential for the next-generation nanotechnology. To this end, it is particularly important to operate AFM devices efficiently via electrical means. Current-induced torques have been theoretically identified as viable driving mechanisms to control the Néel vector in collinear AFM insulators, which involves both field-like (FL) and damping-like (DL) components~\cite{Cheng2014,Gomonay2010,Kamra2018}. Nevertheless, it remains experimentally elusive to properly separate different contributions of the current-induced torques acting on the Néel vector, especially when thermal effects are also present.

Spin-torque ferromagnetic resonance~\cite{liu2011spin,liu2012spin,pai2012spin,mellnik2014spin,macneill2017control} and harmonic Hall analysis~\cite{hayashi2014quantitative,vlietstra2014simultaneous,avci2014interplay,macneill2017thickness,chen2018first,schippers2020large} are two established techniques to quantify current-induced torques in ferromagnetic materials. However, spin-torque AFM resonances are very challenging because the high frequency regime in AFM materials is hard to access. On the other hand, the harmonic Hall analysis utilizes low-frequency drives to probe the spin dynamics and is particularly suitable to elucidate different types of spin torques in magnetic heterostructures. However, the theory of harmonic Hall analysis established based on ferromagnets~\cite{hayashi2014quantitative} is not applicable to AFM materials owing to the distinct N\'{e}el order dynamics~\cite{hals2011phenomenology,Gomonay2010,Cheng2014}, leaving a critical knowledge gap in the rapidly growing frontier of antiferromagnetic spintronics.

In this paper, we fill this knowledge gap by formulating a general theory of the harmonic Hall responses of collinear AFM materials, in which the N\'{e}el vector is driven by current-induced torques arising from an adjacent spin generator. Our findings demonstrate a reliable way to characterize and benchmark the FL torque, the DL torque, as well as concomitant thermal effects in two archetypal AFM insulators, providing an in-time guidance to a series of ongoing experiments of AFM spin dynamics.

As illustrated in Fig.~\ref{fig:AFM/HM_device_1st_harm}, we consider an AFM insulator with easy-plane anisotropy so that the Néel vector is confined in the $x-y$ plane at equilibrium. The magnetic moments can be rotated by a magnetic field. The spin generator is assumed to be a heavy metal (HM) with the spin Hall effect, but in fact, the detailed mechanism of spin generation is not important because the classification of FL and DL torques are purely general. Through the spin Hall effect, an AC current applied to the HM generates AC spin torques that can drive the magnetic moments in the AFM insulator into harmonic oscillations around their equilibrium positions. The variation of N\'eel vector, in turn, changes the spin Hall magnetoresistance (SMR)~\cite{Chen2016review,SMRexperiment,Cheng2016,manchon2017spin} so that the Hall voltage $\tilde{V}_{H}$ is nonlinear in the driving current, creating a series of harmonic Hall signals.

We find that the second harmonic response reflects the FL torque when the magnetic field scans in the $xy$ and the $xz$ planes, while for the $yz$ scan the second harmonic arises solely from the DL torque. In different scans, the second harmonics exhibit a distinct dependence on the field strength, allowing us to separate different contributions of the current-induced torques. In contrast, the first harmonic is only related to the equilibrium direction of the Néel vector independent of current-induced torques. For easy-plane AFM systems with negligible in-plane easy axis anisotropy such as $\alpha-$Fe$_{2}$O$_{3}$, numerical simulations confirm the analytical results of the first and second harmonics, whereas for AFM systems with non-negligible in-plane easy axis anisotropy such as NiO, analytical results are not available thus we only show the numerical results. All numerical calculations are based on material parameters for $\alpha-$Fe$_{2}$O$_{3}$ and NiO. Moreover, thermal contribution to the second harmonics exhibits a distinct angular dependence on the magnetic field compared to that of current-induced torques, allowing us to separate thermal effects from the overall Hall voltage. Our findings suggest that by applying a magnetic field along different directions with varying strength, it is possible to quantify the relative strength of FL and DL torques through harmonic Hall measurements.

\section{Formalism}

\begin{figure}
    \centering
    \includegraphics[width = \linewidth]{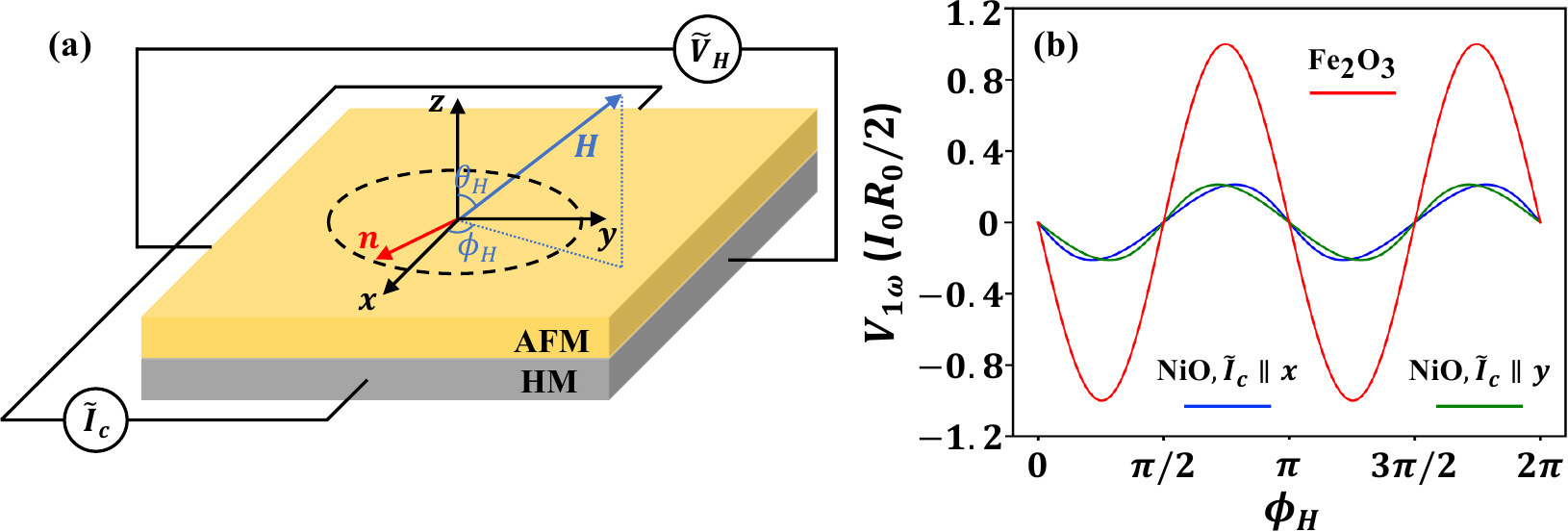}
    \caption{(a) Sample geometry for an AFM/HM harmonic Hall measurement. $\tilde{I}_{c}$ is the AC current, $\tilde{V}_{H}$ is the Hall voltage, $\bm{n}$ is the N\'eel vector, $\bm{H}$ is the external magnetic field, $\theta_{H}$ and $\phi_{H}$ are the polar and azimuthal angles specifying the direction of the applied magnetic field $\bm{H}$ with respect to the direction of current. (b) First harmonic of the Hall voltage in the $xy$ scan for $H=3$T. Red curve: Fe$_{2}$O$_{3}$/HM. Blue and green curves: NiO/HM for $\tilde{I}_c$ parallel and perpendicular to the in-plane easy axis, respectively.}
    \label{fig:AFM/HM_device_1st_harm}
\end{figure}

As sketched in Fig.~\ref{fig:AFM/HM_device_1st_harm}(a), an AFM insulator is in contact with a HM satisfying the spin Hall geometry. An in-plane AC current $\tilde{I}_{c}(t)=I_0\cos\omega t$ generates a spin polarization in the transverse direction that flows in the thickness direction, driving the AFM N\'eel vector $\bm{n}$ through current-induced torques including both DL and FL components. The oscillation of $\bm{n}$ around its equilibrium position results in an oscillating Hall resistance $R_H$, which may include contributions from the SMR and the planar Hall resistance~\cite{Chen2016review,hayashi2014quantitative,baldrati2018full,fischer2018spin,hoogeboom2017negative,lebrun2019anisotropies}. Therefore, the Hall resistance $R_H$ should be a function of $\tilde{I}_{c}(t)$ through the instantaneous $\bm{n}(t)$. In the exchange limit, the N\'{e}el vector $\bm{n}=(\bm{S}_1-\bm{S}_2)/2$ and the magnetization vector $\bm{m}=(\bm{S}_1+\bm{S}_2)/2$ satisfy $|\bm{m}|\ll|\bm{n}|\approx1$, so the contribution of the anomalous Hall effect to $R_H$ can be neglected. By Ohm's law, $R_H$ produces a Hall voltage $\tilde{V}_H = R_{H}\tilde{I_{c}}$ that can be expanded to second order as
\begin{align} \label{eq:resistivity_harmonic}
\tilde{V}_H(t) &= R_H(0) I_{0} \cos \omega t + \frac{1}{2} \frac{\partial R_H(0)}{ \partial \tilde{I}_{c}} I_{0}^{2} (1 + \cos 2\omega t) \notag\\
&= V_0 + V_{1\omega}\cos\omega t + V_{2\omega}\cos2\omega t,
\end{align}
where $R_H(0)$ is the unperturbed Hall resistance ($\tilde{I}_c\rightarrow0$) and $\partial R_H(0) / \partial \tilde{I}_{c}$ is the derivative of $R_H$ with respect to current evaluated at $\tilde{I}_c\rightarrow0$. In Eq.~\eqref{eq:resistivity_harmonic}, $V_0$, $V_{1\omega}$ and $V_{2\omega}$ are the DC and the first two harmonics of the Hall voltage. In the above expansion, $V_0$ only includes the rectification effect, while in real experiments the total DC output may arise from many different mechanisms. This complication, however, does not concern us here because we will focus on the first two harmonics, $V_{1\omega}$ and $V_{2\omega}$, to probes the N\'eel vector configuration.

To determine $\bm{n}$ (hence $R_H$) as a function of the instantaneous current $\tilde{I}_c(t)$, we need to establish the dynamics of $\bm{n}$. In the macrospin approximation, the free energy density of a two-sublattice AFM scaled into the field dimension (in unit of Tesla) can be expressed as
\begin{align}\label{eq:free_energy_spins_Hall}
\frac{E}{\hbar \gamma} =& H_{E} \bm{S}_{1} \cdot \bm{S}_{2} + H_{\perp} \sum_{i=1}^{2} (\bm{S}_{i} \cdot \hat{\bm{z}})^{2} - H_{\parallel} \sum_{i=1}^{2} (\bm{S}_{i} \cdot \hat{\bm{x}})^{2} \notag \\
&- D \hat{\bm{z}} \cdot (\bm{S}_{1} \times \bm{S}_{2}) - \bm{H}_{t} \cdot (\bm{S}_1+\bm{S}_2),
\end{align}
where $\bm{S}_i$ ($i=1,2$) is the unit vector of sublattice magnetic moment, $\gamma$ is the gyromagnetic ratio, and $\hbar$ is the reduced Planck constant. In Eq.~\eqref{eq:free_energy_spins_Hall}, $H_{E}$ is the exchange field between $\bm{S}_{1}$ and $\bm{S}_{2}$, $H_{\perp}$ and $H_{\parallel}$ are the anisotropy fields along the $\bm{z}$ (hard axis) and $\bm{x}$ (easy axis) directions, respectively, $D$ is the Dzyaloshinskii-Moriya (DM) field with the DM vector in the $\bm{z}$ direction, and $\bm{H}_{t} = \bm{H} +\bm{H}_{F}$ is the total magnetic field with $\bm{H}$ being the external field and $\bm{H}_{F}$ arising from the FL torque $\bm{\tau}_{F,i}=\bm{H}_F\times\bm{S}_i$ generated by the AC current. In the following, we will consider two representative AFM systems: (i) $\alpha-$Fe$_{2}$O$_{3}$ (hematite) thin films in which $H_{\parallel}$ is negligible but $D$ and $H_{\perp}$ are both important; (ii) NiO thin films in which $H_{\parallel}$ is non-negligible but $H_{\parallel}\ll H_{\perp}$, and $D=0$. In both cases, $H_{E}$ is several orders of magnitude larger than all other fields (\textit{i.e.}, in the exchange limit), and $|\bm{H}|\gg|\bm{H}_{F}|$ so that the FL effect stemming from $\tilde{I}_c$ is perturbative.

The AFM dynamics can be described by the Landau-Lifshitz-Gilbert-Slonczewski (LLGS) equation
\begin{equation}\label{eq:LLG}
 \frac{d\bm{S}_{i}}{\gamma dt} = \bm{H}_{\rm{eff},i} \times \bm{S}_{i}  +  (\bm{H}_{D}\times\bm{S}_{i})\times\bm{S}_{i},
\end{equation}
where $\bm{H}_{\rm{eff},i} = -(1/\hbar \gamma)\delta E / \delta \bm{S}_{i}$ is the effective field acting on sublattice $\bm{S}_i$ encapsulating the FL effect of $\tilde{I}_c$, and $\bm{H}_{D}$ is the effective field associated with the DL torque $\bm{\tau}_{D,i} = (\bm{H}_{D}\times\bm{S}_{i})\times\bm{S}_{i}$ which cannot be written as a free energy in Eq.~\eqref{eq:free_energy_spins_Hall} due to its non-conservative nature~\footnote{Note that in our convention $\bm{H}_D$ and $\bm{H}_F$ are both defined in the direction of non-equilibrium spin polarization, while in many articles the effective field for DL torque is equivalent to our $(\bm{H}_{D}\times\bm{S}_{i})$.}. Equation~\eqref{eq:LLG} can be equivalently written in terms of the N\'eel vector $\bm{n}$ and the magnetization vector $\bm{m}$~\footnote{The DM interaction breaks the sublattice exchange symmetry, which introduces additional terms not included by the phenomenological theory~\cite{hals2011phenomenology} when rewriting Eq.~\eqref{eq:dynamical_n_Hall} in terms of $\bm{n}$ and $\bm{m}$.}. In the exchange limit, $|\bm{m}|\ll|\bm{n}|\approx1$, we can eliminate $\bm{m}$ using Eq.~\eqref{eq:LLG} and recast the AFM dynamics into a second-order differential equation of $\bm{n}$ as
\begin{align} \label{eq:dynamical_n_Hall}
&\bm{n} \times \left\{  \frac{1}{2H_{E}} \left[ -\frac{\partial^{2} \bm{n}}{\partial t^{2}} - \frac{\partial \bm{n}}{\partial t} \times \bm{H}_{t} + \frac{\partial}{\partial t} \left(\bm{H}_{t} \times \bm{n}\right) \right. \right .\notag  \\
&\qquad\qquad - D\hat{\bm{z}}(D\hat{\bm{z}} \cdot \bm{n}) - D\hat{\bm{z}}\times\bm{H}_{t} - (\bm{n} \cdot \bm{H}_{t})\bm{H}_{t}\bigg] \notag \\ 
& - 2H_{\perp} (\bm{n}\cdot \hat{\bm{z}})\hat{\bm{z}} + 2H_{\parallel} (\bm{n}\cdot \hat{\bm{x}})\hat{\bm{x}} \bigg\} = \bm{n} \times (\bm{n} \times \bm{H}_{D}).
\end{align}
Hereafter, we set $\gamma=1$ and $\hbar=1$ for simplicity unless they need to be reinstated for correct dimensionality. It is worth mentioning that Eq.~\eqref{eq:dynamical_n_Hall} can also be obtained directly from the free energy, known as the nonlinear $\sigma$ model, which is derived in \ref{sec:NLSM}. We emphasize that our theory is applicable to any collinear AFM systems in conjunction with a spin generator, regardless of the microscopic details of how $\bm{H}_F$ and $\bm{H}_D$ are generated.

In harmonic measurements, the frequency of $\tilde{I}_c(t)$ typically ranges from tens of Hz to several kHz, which is far below the intrinsic AFM dynamics lying between GHz and THz~\cite{hayashi2014quantitative}. As a result, the N\'eel vector $\bm{n}$ can be viewed as quasi-static and adiabatically adjusting to the instantaneous driving current $\tilde{I}_c(t)$. Accordingly, we can solve $\bm{n}$ by ignoring all time derivatives in Eq.~\ref{eq:dynamical_n_Hall} (\textit{i.e.}, finding the quasi-equilibrium solution).

\section{Antiferromagnet with negligible $H_{\parallel}$: Case study of Hematite} \label{sec:Fe2O3}

Hematite ($\alpha-$Fe$_{2}$O$_{3}$) is one of the mostly studied easy-plane antiferromangets, where the in-plane anisotropy $H_{\parallel}$ is negligibly small so that the material can be practically viewed as rotationally symmetric around its hard axis. In addition, the strong DM interaction induces a small in-plane magnetic moment $\bm{m}$ perpendicular to $\bm{n}$. As can be deduced from Eq.~\eqref{eq:dynamical_n_Hall} (also see ~\ref{sec:NLSM}), the total magnetic field $\bm{H}_t$ acts effectively as a hard axis anisotropy because of the similarity between $-(\bm{n} \cdot \bm{H}_{t})\bm{H}_{t}$ and $-H_{\perp} (\bm{n}\cdot \hat{\bm{z}})\hat{\bm{z}}$. So quasi-statically, $\bm{n}$ should be perpendicular to $\bm{H}_t$. If $\bm{H}_t$ and the hard-axis $\bm{z}$ are not collinear, they define two non-overlapping easy planes, whose intersection---a line in the $xy$ plane---determines the quasi-equilibrium orientation of $\bm{n}$. This picture holds only when $H_{\parallel}\ll H^2/H_E$ such that the spin-flop threshold is practically zero, which is indeed the case in hematite.

While Eq.(4) does not support an analytical solution in general, what we seek here is the small oscillation of $\bm{n}$ perturbed by the AC current through $\bm{H}_{F}$ and $\bm{H}_{D}$. For this purpose, we define a rotated coordinate system in which $\hat{\bm{x}}_{R} = \bm{n}$ when $\tilde{I}_c=0$ (so $\bm{H}$ lies in the $y_{R}-z_{R}$ plane), $\hat{\bm{z}}_{R} = \hat{\bm{z}}$, and $\hat{\bm{y}}_{R} = \hat{\bm{z}}_{R} \times \hat{\bm{x}}_{R}$. Then, to first order in perturbation, the quasi-equilibrium solution of $\bm{n}$ is given by taking the partial derivatives of Eq.~\eqref{eq:dynamical_n_Hall} with respect to $\bm{H}_{F}$ and $\bm{H}_{D}$. Specifically, taking $d/d\bm{H}_{F}$ of Eq.~\eqref{eq:dynamical_n_Hall} leads to (see \ref{sec:derive5})
\begin{align}\label{eq:df_dH_Hall}
&\mathcal{M} \frac{d \bm{n}}{d \bm{H}_{F}}=\left[ (4H_{E}H_{\perp} + D^{2}) \hat{\bm{y}}_{R} \otimes \hat{\bm{z}}_{R} \right. \notag \\
& \left. +\bm{H} \otimes D\hat{\bm{z}}_{R} - D\hat{\bm{z}}_{R} \otimes \bm{H}- (\hat{\bm{x}}_{R} \times \bm{H}) \otimes \bm{H} \right] \frac{d \bm{n}}{d \bm{H}_{F}} \notag \\
&\qquad = D\hat{\bm{z}}_{R} \otimes \hat{\bm{x}}_{R} + (\hat{\bm{x}}_{R} \times \bm{H}) \otimes \hat{\bm{x}}_{R},
\end{align}
where $\otimes$ denotes the tensor product of two vectors and $\mathcal{M}$ is a $3\times3$ matrix accommodating all tensor products in front of $d\bm{n}/d\bm{H}_{F}$. Because of the constraint $|\bm{n}|^2=1$, $\bm{n}$ has $2$ degrees of freedom, while it has three components $n_x$, $n_y$ and $n_z$. Consequently, $\text{det} \mathcal{M} = 0$ and $\mathcal{M}$ is non-invertible. Now we turn to the vector of spherical coordinates $\bm{\theta} = (\theta, \phi)$ with $\theta$ and $\phi$ the polar and azimuthal angles of $\bm{n}$ (with respect to $\hat{\bm{z}}_{R}$ and $\hat{\bm{x}}_{R}$) in the rotated frame. Using $\bm{\theta}$ removes the redundant degree of freedom. By virtue of the chain rule $d \bm{n} / d \bm{H}_{F} = (d \bm{n} / d \bm{\theta}) \cdot (d \bm{\theta} / d \bm{H}_{F})$, we obtain
\begin{align}\label{eq:dtheta_dH_Hall}
\frac{d \bm{\theta}}{d \bm{H}_{F}} &= \left(\mathcal{M}\frac{d\bm{n}}{d\bm{\theta}}\right)^{-1}_{\rm left} \left[D\hat{\bm{z}}_{R} \otimes \hat{\bm{x}}_{R} + (\hat{\bm{x}}_{R} \times \bm{H}) \otimes \hat{\bm{x}}_{R} \right] \notag\\
&=
\begin{pmatrix}
0 & 0 & 0\\
-\frac1{H_y} & 0 & 0
\end{pmatrix},
\end{align}
where $H_y$ is the projection of the external field onto the $\bm{y}_R$ axis, and $(\cdots)_{\rm left}^{-1}$ is the left-inverse $A^{-1}_{\rm left}\equiv(A^TA)^{-1}A^T$
with $A$ being a full-column rank matrix~\cite{strang2006}. In deriving Eq.~\eqref{eq:dtheta_dH_Hall}, we notice that $d\bm{n}/d\bm{\theta}$ is a $3\times2$ full-column rank matrix, so is $\mathcal{M}d\bm{n}/d\bm{\theta}$. Similarly to Eq.~\eqref{eq:dtheta_dH_Hall}, we obtain
\begin{align}\label{eq:dtheta_dVs_Hall}
\frac{d \bm{\theta}}{d \bm{H}_{D}} &= \frac{2H_{E}}{2(D^{2} + 2H_{E} H_{\perp})H_{y} + D(H^{2} + 4H_{E}H_{\perp} + D^{2})} \notag \\ 
&\times
\begin{pmatrix}
0 & D + H_{y} & H_{z} \\
0 & H_{z} & \frac{DH_{y} + H^{2}_{z} + 4H_{E}H_{\perp} + D^{2}}{H_{y}}
\end{pmatrix}, 
\end{align}
where $H_{z}$ is the projection of the external field onto the $\bm{z}_{R}$ axis.

In the rotated frame, the current-induced oscillation of $\bm{n}$ can be represented by
\begin{align}
 \bm{n} = \hat{\bm{x}}_{R} + \Delta \phi \hat{\bm{y}}_{R} - \Delta \theta \hat{\bm{z}}_{R},
\end{align}
where $\Delta\theta$ and $\Delta\phi$ are the variations of spherical angles driven by the AC current. We assume that $\tilde{I}_c(t)$ is applied in the $\hat{\bm{x}}$ direction, so under the spin Hall geometry, the current-induced fields in the lab frame are $\bm{H}_{F/D} = H_{F/D} \hat{\bm{y}}$ (see footnote 1 again for the direction of effective fields). In the rotated frame,
\begin{subequations}
\begin{align} \label{eq:FL_DL_comoving}
 \bm{H}_{F/D} &= -H_{F/D} \cos \phi_{H} \hat{\bm{x}}_{R} + H_{F/D} \sin \phi_{H} \hat{\bm{y}}_{R},
\end{align}
\end{subequations}
By using Eq.~\eqref{eq:dtheta_dH_Hall} and~\eqref{eq:dtheta_dVs_Hall}, we have
\begin{align}
 \Delta \theta &= \hat{\bm{y}}_{R}\cdot\bm{H}_{D} \left[\frac{d\bm{\theta}}{d\bm{H}_{D}}\right]_{12}, \label{eq:delta_theta} \\
 \Delta \phi &= \hat{\bm{x}}_{R}\cdot\bm{H}_{F} \left[\frac{d \bm{\theta}}{d\bm{H}_{F}}\right]_{21} + \hat{\bm{y}}_{R}\cdot\bm{H}_{D}\left[\frac{d\bm{\theta}}{d\bm{H}_{D}}\right]_{22}. \label{eq:delta_phi}
\end{align}
In the lab frame with spherical coordinates, the external field $\bm{H}=\{\sin\theta_H\cos\phi_H,\sin\theta_H\sin\phi_H,\cos\theta_H\}$ and the N\'eel vector reads
\begin{subequations}\label{eq:nxny}
\begin{align}
 n_{x} &= \sin \phi_{H} + \Delta \phi \cos \phi_{H}, \\
 n_{y} &= -\cos \phi_{H} + \Delta \phi \sin \phi_{H}, \\
 n_{z} &= -\Delta\theta_H,
\end{align}
\end{subequations}
where $\theta_H\in[0,\pi]$ and $\phi_H\in[0,2\pi]$. The Hall resistance arising from SMR and planar Hall effect has the form $R_H = R_{0} n_{x}n_{y}$~\cite{manchon2017spin,baldrati2018full,fischer2018spin,hoogeboom2017negative,lebrun2019anisotropies}. To linear order,
\begin{align}
 n_{x}n_{y} = - \sin(2 \phi_{H})/2 - \Delta \phi \cos (2 \phi_{H}).
\end{align}
By inserting Eq.~\eqref{eq:delta_phi} and~\eqref{eq:nxny} into $R_H$, we finally obtain the harmonic components of Hall voltage as
\begin{subequations} \label{eq:harmonic_hall}
\begin{align}
&V_{1\omega} = -\frac12I_0R_0\sin (2\phi_{H}),  \label{eq:1st_harm} \\
&V_{2\omega} = -\frac12I_0R_0\left[ \frac{H_{F}\cos (2\phi_{H}) \cos \phi_{H}}{H \sin \theta_{H}} \right. \label{eq:2nd_harm}\\
& \left.- \frac{H_{D}H_{E} H \cos \theta_{H} \cos (2\phi_{H}) \sin \phi_{H}}{(D^{2} + 2H_{E}H_{\perp})H \sin \theta_{H} + D(H^{2} + 4H_{E}H_{\perp} + D^{2})/2} \right], \nonumber
\end{align}
\end{subequations}
which is the central result of this paper. Since we have neglected the in-plane easy axis anisotropy $H_{\parallel}$ in deriving Eq.~\eqref{eq:harmonic_hall}, our result becomes invalid if the in-plane component of the external magnetic field, $H \sin{\theta_{H}}$, becomes of order or less than $H_{\parallel}$. In other words, the theory breaks down in the vicinity of $\theta_H=0$ or $\pi$, where $V_{2\omega}$ diverges.

In experiments, it is not realistic to scan the field $\bm{H}$ over arbitrary directions in space. Typically, $\bm{H}$ is rotated and swept on three orthogonal planes, which we take to be the $xy$, $xz$ and $yz$ planes. Within these planes, $V_{2\omega}$ as a function of the field reduces to
\begin{subequations}
\label{eq:harmonic_three_planes_hall}
\begin{align}
&V_{2\omega}^{xy}(\phi_H, H) \sim -\frac{\cos (2\phi_{H}) \cos \phi_{H} }{H } H_{F},  \label{eq:2nd_harm_xy} \\
&V_{2\omega}^{xz}(\theta_H, H) \sim -\frac{1}{H \sin \theta_{H}} H_{F}, \label{eq:2nd_harm_xz} \\
&V_{2\omega}^{yz}(\theta_H, H) \sim  \label{eq:2nd_harm_yz}\\
&\frac{-H_{E} H \cos \theta_{H} H_{D}}{H(D^{2} + 2H_{E}H_{\perp})\sin \theta_{H} + \frac{D}{2}(H^{2} + 4H_{E}H_{\perp} + D^{2})},\notag
\end{align}
\end{subequations}
where $V_{2\omega}^{xy}$ has $\theta_{H} = \pi/2$, $V_{2\omega}^{xz}$ has $\phi_{H} = 0$ and $V_{2\omega}^{yz}$ has $\phi_{H} = \pi / 2$. We can see from Eq.~\eqref{eq:harmonic_three_planes_hall} that $V_{2\omega}^{xy}$ and $V_{2\omega}^{xz}$ only depends on the FL torque while $V_{2\omega}^{yz}$ only stems from the DL torque. This fact indicates that we are able to separate different current-induced torques by comparing these field scans.

\begin{figure}
    \centering
    \includegraphics[width = \linewidth]{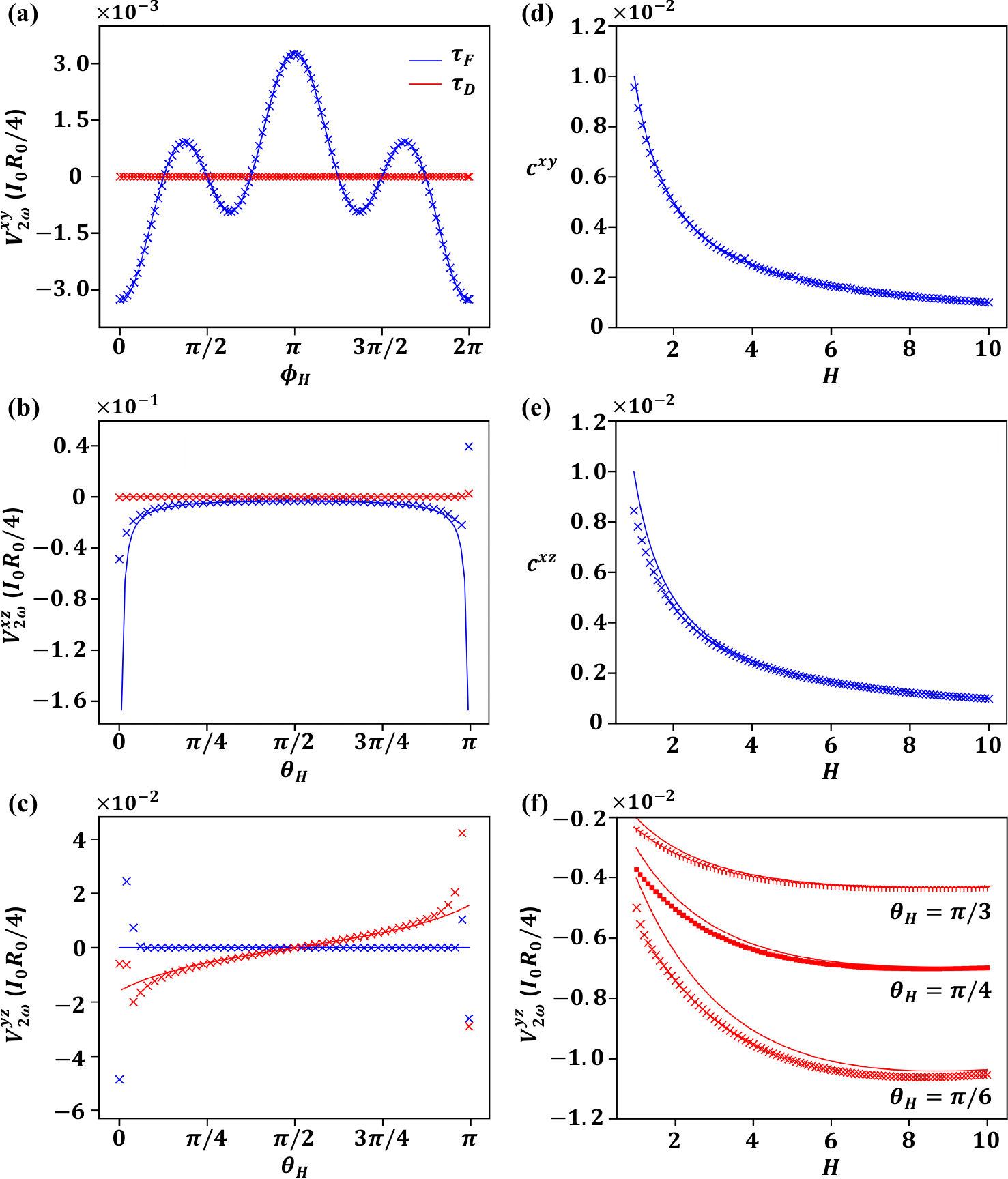}
    \caption{Second harmonic Hall signals of a hematite/HM heterostructure in (a) $xy$ scan, (b) $xz$ scan, and (c) $yz$ scan driven by either the FL torque (blue) or the DL torque (red) at $H=3$T. (d) and (e) show the field dependence of $V_{2\omega}^{xy}$ and $V_{2\omega}^{xz}$ fitted by $V_{2\omega}^{xy} = -c^{xy} \cos (2\phi_{H}) \cos \phi_{H}$ and $V_{2\omega}^{xz} = -c^{xz}/\sin \theta_{H}$, respectively. (f) shows the field dependence in the $xz$ scan for three different $\theta_H$. In all sub-figures, crossed dots are from numerical simulations while solid curves are from theory. Units: magnetic field $H$ in Tesla; all $V_{2\omega}$ components and fitting coefficients are normalized to $I_{0}R_{0}/4$.}
    \label{fig:Fe2O3_second_harm}
\end{figure}

To verify the theoretical results, we numerically solve the LLGS equation~\eqref{eq:LLG} with varying direction and strength of the magnetic field. We use material parameters of $\alpha-$Fe$_{2}$O$_{3}$~\cite{besser1967magnetocrystalline,lebrun2019anisotropies}: $H_{E} = 900$T, $H_{\perp} = 0.02$T, $H_{\parallel} = 10^{-4}$T, $D = 2$T, and take $\alpha = 0.01$ (a typical value). To keep the oscillation of $\bm{n}$ characterized by Eq.~\eqref{eq:delta_theta} and~\eqref{eq:delta_phi} linear in the current-induced fields $H_{F}$ and $H_{D}$ (accordingly, $V_{2\omega}$ is quadratic in $I_0$), the perturbation cannot be too large. Close to the breaking point of linear response regime, we take $H_{F} = 5 \times 10^{-3}$T and $H_{D} = 2.5 \times 10^{-4}$T, which typically corresponds to a current density of $10^{6}\sim 10^{7} \text{A}/\text{cm}^{2}$~\cite{egecan2021quantifying}. The numerical result of the first harmonic is shown in Fig.~\ref{fig:AFM/HM_device_1st_harm}(b) (red curve), which exhibits a perfect $-\sin (2\phi_{H})$ shape, agreeing very well with Eq.~\eqref{eq:1st_harm}. Figure~\ref{fig:Fe2O3_second_harm} (a)-(c) plot the numerical simulations for $V_{2\omega}^{xy}$, $V_{2\omega}^{xz}$ and $V_{2\omega}^{yz}$, respectively, where each curve includes contribution from either the FL torque or the DL torque so that their physical consequences can be clearly compared. We observe that the FL torque (blue curves) $\bm{\tau}_{F,i}=\bm{H}_F\times\bm{S}_i$ only affects the $xy$ and $xz$ scans while the DL torque (red curves) $\bm{\tau}_{D,i}=(\bm{H}_D\times\bm{S}_i)\times\bm{S}_i$ only manifests in the $yz$ scan except for $\theta_H$ very close to $0$ and $\pi$ where the in-plane component of $\bm{H}$ is too small to dictate the direction of $\bm{n}$. These results are consistent with Eq.~\eqref{eq:harmonic_three_planes_hall}.

Besides the angular dependence, we also plot the field dependence of the second harmonics for the three special scans in Fig.~\ref{fig:Fe2O3_second_harm}(d)-(f). For a set of discrete $H$ ranging from $0-10$T, we fit the numerical results of $V_{2\omega}^{xy}$ and $V_{2\omega}^{xz}$ by Eq.~\eqref{eq:2nd_harm_xy} and~\eqref{eq:2nd_harm_xz}, which are plotted by solid curves in Fig.~\ref{fig:Fe2O3_second_harm}(d) and (e), where they decrease monotonically as $1/H$. The numerical simulation agrees well with theory at large fields. At small fields, the deviations become visible but remain minor. The field dependence of $V_{2\omega}^{yz}$ is complicated according to Eq.~\eqref{eq:2nd_harm_yz}, so we choose three particular values of $\theta_H$ and directly plot Eq.~\eqref{eq:2nd_harm_yz} with the same material parameters in Fig.~\ref{fig:Fe2O3_second_harm}(f). Again, we see good agreement between numerical and theoretical results at large fields. Also, the agreement becomes increasingly good for larger $\theta_H$. The deviation at small fields shown in Fig.~\ref{fig:Fe2O3_second_harm}(d)-(f) is a common feature for all three scans, which, as we discussed above, can be attributed to the vanishing in-plane projection of $\bm{H}$ that is not sufficiently strong to anchor the direction of $\bm{n}$. Furthermore, we observe a fine structure in Fig.~\ref{fig:Fe2O3_second_harm}: all curves are actually non-monotonic that they slightly curve up at extremely large fields. This behavior, being almost invisible, can be traced back to Eq.~\eqref{eq:2nd_harm_yz}, which has a form of $H/(H^{2} + a H + b)$ with $a$ and $b$ positive constants.

In order to understand why different components of the current-induced torques manifest in different field scans, we provide an anatomy of spin reactions to the current-induced torques. As mentioned above, a magnetic field acts effectively as a hard axis anisotropy on the N\'eel vector so that $\bm{n}(t)$ is perpendicular to both $\bm{H}_t(t)=\bm{H}+\bm{H}_F(t)$ and the hard axis $\bm{z}$. When $\bm{H}$ varies on the $yz$ plane, an oscillating $\bm{H}_F(t)$ along $\bm{y}$ maintains $\bm{H}_t(t)\cdot\bm{n}(t)=0$ at all times, hence no torque acting on $\bm{n}$. This is why the FL torque does not show up in the $yz$ scan. If $\bm{H}$ moves away from the $yz$ plane, however, an oscillating $\bm{H}_F$ along $\bm{y}$ will generate an oscillating energy perturbation, leading to an oscillation of the intersection between the easy plane ($xy$ plane) and the plane normal to $\bm{H}_t$. Accordingly, $\bm{n}(t)$ must oscillate with this perturbation to stay on the intersection at all times. This explains why the FL torque affects both $xy$ and $xz$ scans. To study the DL torque, we move to the rotated frame in which the unperturbed $\bm{n}$ is along $\bm{x}_R$ while $\bm{H}$ is along $\bm{y}_R$. Then $\bm{H}_{D}$ can be projected onto $\bm{x}_{R}$ and $\bm{y}_{R}$ axes. If $\bm{H}_{D} \cdot \hat{\bm{y}}_{R} \neq 0$, both $\bm{S}_{i}$ will be canted towards $\bm{y}_{R}$ under the DL torque, developing a slight non-collinearity between $\bm{S}_1$ and $\bm{S}_2$. This non-collinearity will subsequently initiate the exchange torque which rotates $\bm{S}_1$ and $\bm{S}_2$ towards out-of-plane directions. Accordingly, the N\'eel vector also develops an out-of-plane tilting $\Delta n_{z}$, namely $\bm{n} = \hat{\bm{x}}_{R} + \Delta n_{z} \hat{\bm{z}}_{R}$. Because $\bm{H} \parallel \hat{\bm{y}}_{R}$ and $\bm{n}$ lies in the $x_{R}$-$z_{R}$ plane, we have $\bm{n}\cdot\bm{H} = 0$ thus no oscillation in the $xy$ plane will be induced. This is why the DL torque does not exhibit in the $xy$ scan. However, if $\bm{H}$ has an out-of-plane component, $\bm{n} = \hat{\bm{x}}_{R} + \Delta n_{z} \hat{\bm{z}}_{R}$ will no more be perpendicular to $\bm{H}$ (\textit{i.e.}, $\bm{n}\cdot \bm{H} \neq 0$), which must be followed by a rotation in the $x_{R}$-$y_{R}$ plane to accommodate the energy penalty brought by $\Delta n_{z}$. This explains why the DL torque manifests in the $yz$ scan. From the above analysis, we see that $\bm{H}_{D} \cdot \hat{\bm{y}}_{R} \neq 0$ and $\bm{H} \cdot \hat{\bm{z}}_{R} \neq 0$ are two necessary conditions for $\bm{H}_D$ to play a rule in $V_{2\omega}$. Thus, only the $yz$ scan satisfies both conditions, whereas $\bm{H} \cdot \hat{\bm{z}}_{R} = 0$ in the $xy$ scan and $\bm{H}_{D} \cdot \hat{\bm{y}}_{R} = 0$ in the $xz$ scan. 

The current-induced torques are inevitably accompanied by Joule heating, which generates a temperature gradient along the $\bm{z}$ direction. This temperature gradient can drive a spin current via the spin Seebeck effect (SSE) and inject spins into the HM, converting into a transverse voltage through the inverse spin Hall effect. Since the heating effect is proportional to $\tilde{I}_{c}^{2}$, the thermal contribution will be mixed with the second harmonic signal. As discussed above, in hematite with negligible in-plane anisotropy, the N\'eel vector $\bm{n}$ is perpendicular to $\bm{H}$ such that the AFM material is effectively in the spin-flop phase immediately when $\bm{H}$ is being applied. Therefore, the magnon excitations carry non-equilibrium spins against the small magnetization along the magnetic field~\cite{li2020spin,reitz2020spin}. Under the spin Hall geometry, the SSE-induced Hall voltage is in the direction of $\hat{\bm{z}}\times\bm{H}$, which, when being projected onto the $\bm{y}$ axis (see Fig.~\ref{fig:AFM/HM_device_1st_harm}), produces a $\cos\phi_H$ dependence in $V_{2\omega}^{xy}$ and a $\sin\theta_H$ dependence in $V_{2\omega}^{xz}$. Therefore, the thermal contribution can be separated thanks to its different angular dependence from Eq.~\eqref{eq:harmonic_three_planes_hall}. For the $yz$ scan, the thermally driven spin current does not contribute to $\tilde{V}_H$ under the spin Hall geometry.

To close the discussion of this section, we mention that our theory applies only to collinear AFM materials in the exchange limit. In synthetic AFM systems~\cite{duine2018synthetic} where $|\bm{m}|\ll|\bm{n}|$ may not be true or in non-collinear AFM materials whose order parameter are not simply $\bm{n}$~\cite{yamane2019dynamics,shukla2022spin}, our theory becomes invalid. Moreover, since the forms of SMR and planar Hall resistance in non-collinear systems are distinct from their collinear counterparts~\cite{dong2017spin}, a different formalism is needed to study the harmonic Hall responses of non-collinear AFM materials. In addition, if the heavy metal spin generator is replaced by transition metal dichalcogenides such as WTe$_{2}$~\cite{macneill2017control}, an out-of-plane spin polarization can be generated by an in-plane current, which breaks the simple spin Hall geometry~\cite{tserkovnyak2014spin}, and our theory must be modified accordingly.

\section{Antiferromagnet with non-negligible $H_{\parallel}$: case study of $\text{NiO}$}
\label{sec:NiO}

Nickel oxide (NiO) is a representative collinear AFM insulator whose physical properties are well understood. The dynamics of NiO is affected by a strong hard-axis anisotropy $K_{\perp}$ [along $(111)$] and a relatively weak but non-negligible easy-axis anisotropy $K_{\parallel}$ [along $(11\bar{2})$], while the DM interaction vanishes identically. In the presence of both $K_{\perp}$ and $K_{\parallel}$, however, the equilibrium orientation of $\bm{n}$ cannot be obtained analytically as a function of the field direction. Consequently, the oscillation of $\bm{n}$ around its equilibrium can only be studied by numerical simulations.

A NiO thin film is typically grown in the (111) direction so that the applied AC current $\tilde{I}_{c}(t)$ still lies in the easy plane. However, we need to consider two distinct geometry due to the existence of $K_{\parallel}$, namely, $\tilde{I}_{c}(t)$ parallel and perpendicular to the in-plane easy axis (\textit{i.e.}, $\bm{x}$ axis). To describe the field direction consistently, we redefine its azimuthal angle $\phi_{H}$ with respect to the direction of the current, and accordingly, we associate the $xz$ ($yz$) scan with $\phi_{H} = 0$ ($\phi_{H} = \pi / 2$). We choose material parameters of NiO to be~\cite{rezende2019introduction}: $H_{E} = 968.4$T, $H_{\perp} = 0.635$T, $H_{\parallel} = 0.011$T, $D = 0$T, $\alpha = 0.01$, and as in the preceding section, $H_{F} = 5 \times 10^{-3}$T and $H_{D} = 2.5 \times 10^{-4}$T. In contrast to hematite in which the spin-flop threshold is practically zero, the spin-flop field in NiO is as large as $\sqrt{H_EH_{\parallel}}\sim4.6$T. In the following, we restrict the field strength to be below $4$T so that the system does not undergo the spin-flop transition.

\begin{figure}
    \centering
    \includegraphics[width = \linewidth]{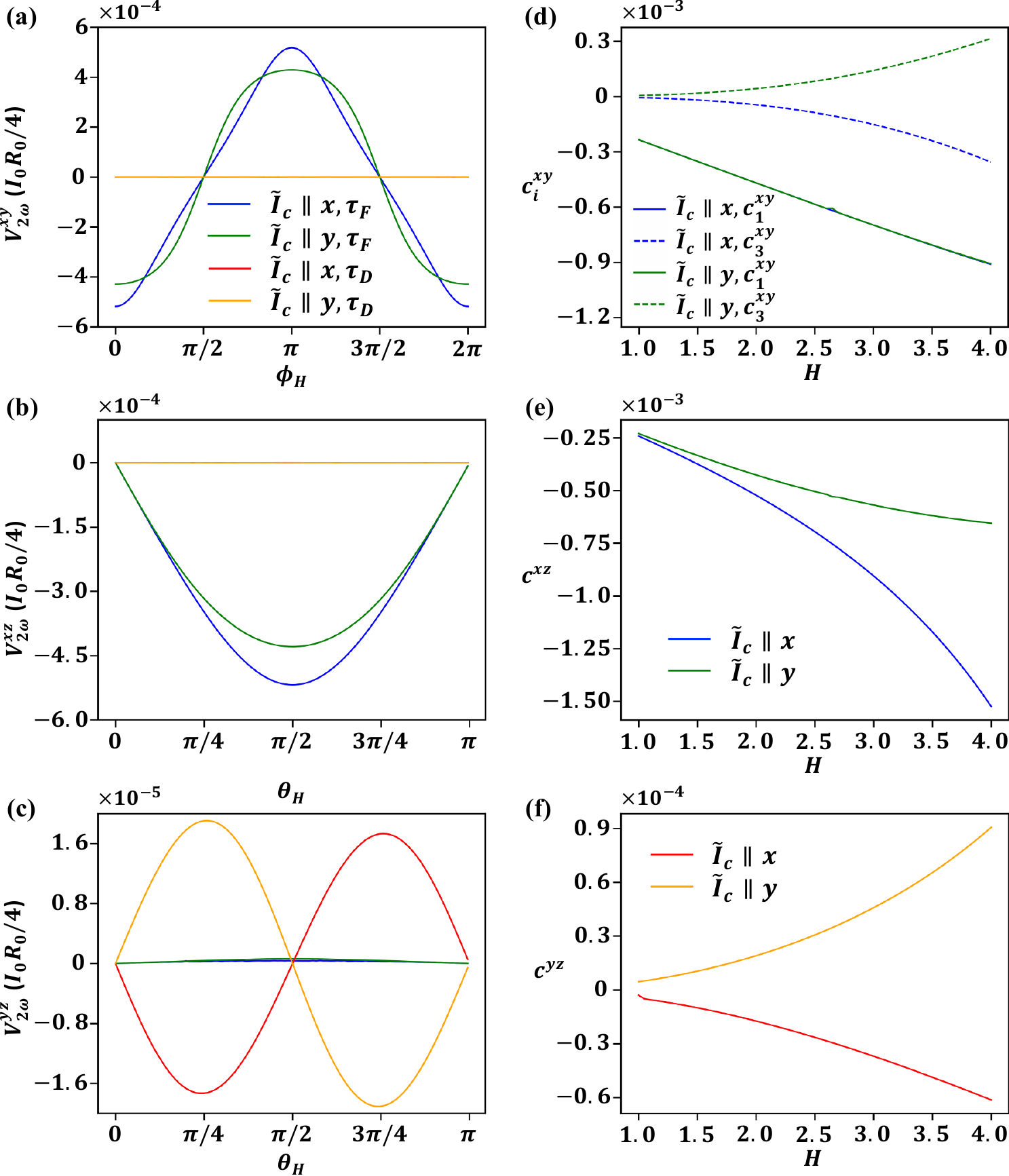}
    \caption{Second harmonic Hall signals of a NiO/HM heterostructure in (a) $xy$ scan, (b) $xz$ scan, and (c) $yz$ scan driven by either the FL torque $\bm{\tau}_F$ or the DL torque $\bm{\tau}_D$ for both $\tilde{I}_c\parallel\bm{x}$ and $\tilde{I}_c\parallel\bm{y}$ at $H=3$T. (d), (e) and (f) plot the fitting coefficients in the corresponding scans defined as $V_{2\omega}^{xy} = c^{xy}_{1} \cos \phi_{H} + c^{xy}_{3} \cos (3 \phi_{H})$, $V_{2\omega}^{xz} = c^{xz} \sin \theta_{H}$, and $V_{2\omega}^{yz} = c^{yz} \sin 2\theta_{H}$. Color schemes for each curve are explained in the legends. Units: magnetic field $H$ in Tesla; all $V_{2\omega}$ components and fitting coefficients are normalized to $I_{0}R_{0}/4$.}
    \label{fig:NiO_second_harm}
\end{figure}

Figure~\ref{fig:AFM/HM_device_1st_harm}(b) (blue and green curves) plots $V_{1\omega}$ in the $xy$ scan for $H=3$T. The results of $\tilde{I}_c\parallel\bm{x}$ and $\tilde{I}_c\parallel\bm{y}$ largely overlap and essentially follows the same pattern $-\sin (2\phi_{H})$ as that in hematite with a substantially reduced amplitude. The sinusoidal pattern agrees well with recent measurements~\cite{hoogeboom2017negative}. The minor difference between the two curves is attributed to the existence of $H_{\parallel}$, which becomes larger for smaller field strength.

Figure~\ref{fig:NiO_second_harm}(a)--(c) plot the second harmonics, $V_{2\omega}^{xy}$, $V_{2\omega}^{xz}$ and $V_{2\omega}^{yz}$ (with respect to the $\tilde{I}_c$ direction), respectively. Similar to the previous section, each curve includes either the FL or the DL torque but not both in order to clearly separate their contributions. It turns out that, again, the FL torque manifests in the $xy$ and $xz$ scans whereas the DL torque only affects the $yz$ scan. We fit the numerical results in (a)--(c) for both $\tilde{I}_c\parallel\bm{x}$ and $\tilde{I}_c\parallel\bm{y}$ by the following functions:
\begin{subequations}
\begin{align}
 V_{2\omega}^{xy} &= c^{xy}_{1} \cos \phi_{H} + c^{xy}_{3} \cos3\phi_{H}, \\
 V_{2\omega}^{zy} &= c^{zy} \sin \theta_{H}, \\
 V_{2\omega}^{yz} &= c^{yz} \sin2\theta_{H},
\end{align}
\end{subequations}
and the fitting results are plotted in Fig.~\ref{fig:NiO_second_harm}(d)--(f). In the $xy$ scan, $c^{xy}_1$ is the same while $c^{xy}_3$ is opposite for $\tilde{I}_c\parallel\bm{x}$ and $\tilde{I}_c\parallel\bm{y}$. In the $xz$ scan $c^{yz}$ is of the same sign for both current directions, while in the $yz$ scan $c^{yz}$ is opposite for different current directions. In all cases, the amplitude of fitting coefficients increase monotonically with an increasing magnetic field. 

To further understand the impact of in-plane anisotropy $H_{\parallel}$, we fix $H_{\perp}$ and gradually increase $H_{\parallel}$ from $0$. During this virtual process, we observe that the three scans shown in Fig.~\ref{fig:Fe2O3_second_harm}(a)-(c) smoothly transition into those shown in Fig.~\ref{fig:NiO_second_harm}(a)-(c) until the curves no long change shapes, while the overall amplitudes of all second harmonic components continue to decrease. The latter pattern indicates that easy-axis anisotropy inhibits the oscillation of $\bm{n}$, hence suppressing the harmonic signals. In contrast, by fixing $H_{\parallel}$ while increasing $H_{\perp}$ from $0$, the angular dependence remains almost the same fo  $V_{2\omega}^{xy}$ and $V_{2\omega}^{xz}$ while the amplitude of $V_{2\omega}^{yz}$ decreases with a rather minor change in shape. AFM materials with more complicated form of anisotropy may bring about different harmonic responses, which goes beyond the scope of this paper and will be left for future studies.

\section{Conclusion}

In summary, we formulated a general theory of the harmonic Hall responses of collinear AFM materials driven by current-induced torques. As two model systems, we studied $\alpha-$Fe$_{2}$O$_{3}$ (hematite) and NiO, where the difference in magnetic anisotropy and the DM interaction leads to distinct behavior. In hematite, we obtained analytical results of the second harmonics for field scanning in the $xy$, $xz$ and $yz$ planes, which agrees very well with numerical simulations. In NiO, the presence of easy-axis anisotropy complicates the problem and analytical results are not available. Instead, we investigated the second harmonics fully numerically, acquiring a parallel understanding of each scan with hematite. The angular and field dependence of various components in the second harmonics make it possible to not only unambiguously distinguish the field-like and damping-like components but also separate thermal effects from the current-induced torques.

Our theory enables a quantitative harmonic analysis of the Hall response in AFM materials with easy-plane anisotropy, providing direct guidance to ongoing and future experiments. We anticipate further generalizations of our findings to incorporate more complicated device geometries such as out-of-plane damping-like torques recently claimed in WTe$_2$ and AFM materials with more complicated magnetic anisotropy.

\section{Acknowledgement}
We acknowledge inspiring discussions with E. Cogulu, A. D. Kent, N. Statuto, and Y. Cheng. This work was motivated by the experiment reported in Ref.~\cite{egecan2021quantifying}, where the theoretical analysis is based on and be a special case of the general findings presented here. This work is supported by the Air Force Office of Scientific Research under grant FA9550-19-1-0307.

\appendix
\section{Alternative derivation of Eq.~\eqref{eq:dynamical_n_Hall}: Nonlinear $\sigma$ model}
\label{sec:NLSM}
Using the definitions of $\bm{m}$ and $\bm{n}$, we can recast the free energy in the form of
\begin{align}\label{eq:free_energy_nm_Hall}
E =& 2H_{E} \bm{m}^{2} + 2H_{\perp} (\bm{n} \cdot \hat{\bm{z}})^{2} - 2H_{\parallel} (\bm{n} \cdot \hat{\bm{x}})^{2} \notag \\
&\quad - 2D \hat{\bm{z}} \cdot (\bm{n} \times \bm{m}) - 2\bm{H}_{t} \cdot \bm{m},
\end{align}
where we set $\hbar = 1$ and $\gamma = 1$ as in the main text. The corresponding Lagrangian can then be written as~\cite{cheng2014aspects}
\begin{equation}\label{eq:lagrangian_nm_Hall}
\mathcal{L} = -2 \bm{m} \cdot \left(\bm{n} \times \frac{\partial \bm{n}}{\partial t} \right) - E - \lambda \bm{m} \cdot \bm{n} - \mu \left(\bm{n}^2 -1 \right),
\end{equation}
where the first term is the Wess-Zumino term arising from the spin Berry phase of each sublattice, and $\lambda$ and $\mu$ are Lagrange multipliers enforcing the constraints of $\bm{n} \cdot \bm{m} = 0$ and $\bm{n}^{2} = 1$. In the path integral of the system, we can integrate out the variable $\bm{m}$ to extract the effective Lagrangian in terms of $\bm{n}$ alone as
\begin{align}
 Z &= \int \mathcal{D}\bm{n} \mathcal{D}\bm{m} \mathcal{D}\lambda \mathcal{D}\mu \exp \left[ i \int dt \mathcal{L}(\bm{m},\bm{n})\right] \notag\\
 &=\int\mathcal{D}\bm{n}\mathcal{D}\lambda \mathcal{D}\mu \exp\left[i\int dt\mathcal{L}_{\rm eff}(\bm{n})\right],
\end{align}
where
\begin{align}\label{eq:lagrangian_n_Hall}
&\mathcal{L}_{\rm eff}(\bm{n}) = \frac{1}{2H_{E}} \left( -\bm{n} \times \frac{\partial \bm{n}}{\partial t} + \bm{H}_{t} + D \hat{\bm{z}} \times \bm{n} - \frac{\lambda}{2} \bm{n} \right)^{2} \notag \\
&\quad - 2H_{\perp} (\bm{n} \cdot \hat{\bm{z}})^{2} + 2H_{\parallel} (\bm{n} \cdot \hat{\bm{x}})^{2} - \mu \left(\bm{n}^2 -1 \right),
\end{align}
where a constant term generated by the integration is omitted. Using the Euler-Lagrange equations for $\lambda$ and $\mu$, we obtain $\lambda = 2 \bm{n} \cdot \bm{H}_{t}$ and $\bm{n}^{2} = 1$. Then, substituting $\lambda = 2 \bm{n} \cdot \bm{H}_{t}$ back into the Euler-Lagrange equation for $\bm{n}$, we obtain
\begin{align} \label{eq:EL_n}
& \frac{1}{2H_{E}} \left\{ \bm{n} \left( \frac{\partial  \bm{n}}{\partial t}\right)^{2} - \frac{\partial^{2} \bm{n}}{\partial t^{2}} - \frac{\partial \bm{n}}{\partial t} \times \bm{H}_{t} +  \frac{\partial }{\partial t} \left(\bm{H}_{t} \times \bm{n}\right)   \right. \notag \\
&\quad + D^{2} \bm{n} +3\left(D\hat{\bm{z}} \cdot \frac{\partial \bm{n}}{\partial t}\right) \bm{n}  - (D\hat{\bm{z}} \cdot \bm{n})D\hat{\bm{z}}  \notag\\ 
&\quad \left. + (\bm{n} \cdot \bm{H}_{t})^{2} \bm{n} + \bm{H}_{t}\times D\hat{\bm{z}} - (\bm{n}\cdot\bm{H}_{t}) \bm{H}_{t} \right\}   \notag \\
&\quad - 2H_{\perp} (\bm{n}\cdot \hat{\bm{z}})\hat{\bm{z}} + 2H_{\parallel} (\bm{n}\cdot \hat{\bm{x}})\hat{\bm{x}} = \mu \bm{n}.
\end{align}
To eliminate $\mu$, we take the cross product of Eq.~\eqref{eq:EL_n} with $\bm{n}$, ending up with
\begin{align} \label{eq:dynamical_n_Hall_suppl}
&\bm{n} \times \left\{  \frac{1}{2H_{E}} \left[ -\frac{\partial^{2} \bm{n}}{\partial t^{2}} - \frac{\partial \bm{n}}{\partial t} \times \bm{H}_{t} + \frac{\partial}{\partial t} \left(\bm{H}_{t} \times \bm{n}\right) \right. \right .\notag  \\
&\quad \left. \left. - (D\hat{\bm{z}} \cdot \bm{n})D\hat{\bm{z}} + \bm{H}_{t} \times D\hat{\bm{z}} - (\bm{n} \cdot \bm{H}_{t})\bm{H}_{t} \right] \right.\notag \\ 
&\quad \left. - 2H_{\perp} (\bm{n}\cdot \hat{\bm{z}})\hat{\bm{z}} + 2H_{\parallel} (\bm{n}\cdot \hat{\bm{x}})\hat{\bm{x}} \right\} = 0,
\end{align}
which is consistent with Eq.~\eqref{eq:dynamical_n_Hall} except for the absence of the DL torque. This is because the DL torque cannot be reproduced by the energy alone; it must be added as a damping term using the Rayleigh dissipation function, the form of which is chosen from the LLGS equation. It should be noted that in Eq.~\eqref{eq:dynamical_n_Hall_suppl}, the magnetic field acts effectively as a hard axis anisotropy since $- (\bm{n} \cdot \bm{H}_{t})\bm{H}_{t}$ and $- H_{\perp} (\bm{n}\cdot \hat{\bm{z}})\hat{\bm{z}}$ share the same form.

\section{Detailed derivation of Eq.~\eqref{eq:df_dH_Hall}}
\label{sec:derive5}

Under the chain rule $d / d\bm{H}_{F} = (d / d\bm{H}_{t})(d \bm{H}_{t} / d\bm{H}_{F}) = d / d\bm{H}_{t}$, taking $d/d\bm{H}_{t}$ on Eq.~\eqref{eq:dynamical_n_Hall} yields
\begin{align} \label{eq:df_over_dH_suppl}
& \frac{1}{2 H_{E}} \left\{  - (\bm{n} \cdot \bm{H}_{t}) \left( \bm{n} \times \mathcal{\bm{I}}  - \bm{H}_{t} \times \frac{d \bm{n}}{d \bm{H}_{t}}  \right) + (\bm{n} \cdot D\hat{\bm{z}}) \mathcal{\bm{I}} \right. \notag  \\
& + \bm{H}_{t} \otimes \left( D\hat{\bm{z}} \cdot \frac{d \bm{n}}{d \bm{H}_{t}}   \right)  - D\hat{\bm{z}} \otimes \left( \bm{H}_{t} \cdot \frac{d \bm{n}}{d \bm{H}_{t}}  \right)  \notag \\
& - D\hat{\bm{z}} \otimes \bm{n} -(\bm{n} \times \bm{H}_{t}) \otimes \left(\bm{H}_{t} \cdot \frac{d \bm{n}}{d \bm{H}_{t}} + \bm{n}  \right) \notag \\
&\left. -D^{2} \left[ (\bm{n} \times \hat{\bm{z}}) \otimes \left(  \hat{\bm{z}} \cdot \frac{d \bm{n}}{d \bm{H}_{t}} \right) - (\bm{n} \cdot \hat{\bm{z}}) \left( \hat{\bm{z}} \times \frac{d \bm{n}}{d \bm{H}_{t}}  \right) \right] \right\} \notag \\
&-2H_{\perp} \left[ (\bm{n} \times \hat{\bm{z}}) \otimes \left( \hat{\bm{z}} \cdot \frac{d \bm{n}}{d \bm{H}_{t}} \right)  - (\bm{n} \cdot \hat{\bm{z}}) \left( \hat{\bm{z}} \times \frac{d \bm{n}}{d \bm{H}_{t}}  \right)\right] \notag \\
&+2H_{\parallel} \left[ (\bm{n} \times \hat{\bm{x}}) \otimes \left( \hat{\bm{x}} \cdot \frac{d \bm{n}}{d \bm{H}_{t}} \right)  - (\bm{n} \cdot \hat{\bm{x}}) \left( \hat{\bm{x}} \times \frac{d \bm{n}}{d \bm{H}_{t}}  \right)\right] \notag \\
& -\bm{n} \otimes \left( \bm{H}_{D} \cdot \frac{d \bm{n}}{d \bm{H}_{t}} \right) - (\bm{n} \cdot \bm{H}_{D}) \frac{d \bm{n}}{d \bm{H}_{t}} = 0,
\end{align}
where terms like $\hat{\bm{z}} \times (d\bm{n} / d\bm{H}_{t})$ should be understood as $
[\hat{\bm{z}} \times (d\bm{n} / d\bm{H}_{t})]_{ij} = \epsilon_{iab} \hat{\bm{z}}_{a} (d\bm{n}_{b} / d \bm{H}_{t,j})$, and terms like $\hat{\bm{z}} \cdot (d \bm{n}/d \bm{H}_{t})$ are defined as $[\hat{\bm{z}} \cdot (d \bm{n}/d \bm{H}_{t})]_{i} = \hat{\bm{z}}_{j} (d \bm{n}_{j} / d \bm{H}_{t,i})$ with Einstein summation rule assumed, and $\mathcal{\bm{I}}$ is the identity matrix. Eq.~\eqref{eq:df_over_dH_suppl} is general and applicable to both types of AFM systems discussed in Sec.~\ref{sec:Fe2O3} and~\ref{sec:NiO}.
In the linear response regime, Eq.~\eqref{eq:df_over_dH_suppl} should be evaluated at zero current (\textit{i.e.}, $\bm{H}_{F} = \bm{H}_{D} = 0$). In this case, $\bm{H}_{t} = \bm{H}$, and as mentioned in the main text, we can ignore $H_{\parallel}$ in hematite so that $\bm{n} = \hat{\bm{x}}_{R}$ in the rotated frame and $\bm{n} \cdot \bm{H} = 0$. Using these relations, we are able to simplify Eq.~\eqref{eq:df_over_dH_suppl} into
\begin{align} \label{eq:df_over_dH_simplify_suppl}
& \frac{1}{2 H_{E}} \left\{ D^{2} \hat{\bm{y}}_{R} \otimes \left(  \hat{\bm{z}}_{R} \cdot \frac{d \bm{n}}{d \bm{H}_{t}} \right) + \bm{H} \otimes \left( D\hat{\bm{z}}_{R} \cdot \frac{d \bm{n}}{d \bm{H}_{t}}   \right)  \right. \notag  \\
& - D\hat{\bm{z}}_{R} \otimes \left( \bm{H} \cdot \frac{d \bm{n}}{d \bm{H}_{t}}  \right)  -(\hat{\bm{x}}_{R} \times \bm{H}) \otimes \left(\bm{H} \cdot \frac{d \bm{n}}{d \bm{H}_{t}} + \hat{\bm{x}}_{R}  \right) \notag \\
&\left. - D\hat{\bm{z}}_{R} \otimes \hat{\bm{x}}_{R} \right\} +2H_{\perp} \hat{\bm{y}}_{R} \otimes \left( \hat{\bm{z}}_{R} \cdot \frac{d \bm{n}}{d \bm{H}_{t}} \right) = 0.
\end{align}
By invoking the identity $\hat{\bm{y}}_{R} \otimes [\hat{\bm{z}}_{R} \cdot ( d\bm{n} / d\bm{H}_{t}) ]= (\hat{\bm{y}}_{R} \otimes \hat{\bm{z}}_{R}) (d\bm{n} / d\bm{H}_{t})$, we finally arrive at Eq.~\eqref{eq:df_dH_Hall}.

\bibliographystyle{elsarticle-num} 
\bibliography{Reference_final}

\end{document}